\documentclass[useAMS,usenatbib]{mn2e}
\def\msolar {$M_\odot$}
\def\hii{H{\sc ii}}
\def\msun {$M_\odot$}

\title[The star-forming content of the W\,3 GMC]{The star-forming content of 
the W\,3 giant molecular cloud}
\author[T. J. T. Moore, et al.]{T. J. T. Moore$^{1}$\thanks{E-mail:
tjtm@astro.livjm.ac.uk}, D. E. Bretherton$^{1}$, T. Fujiyoshi$^{2}$, 
N. A. Ridge$^{3}$, J. Allsopp$^{1}$, \and M. G. Hoare$^{4}$, 
S. L. Lumsden$^{4}$ and J. S. Richer$^{5}$\\
$^{1}$Astrophysics Research Institute, Liverpool John Moores
  University, Twelve Quays House, Egerton Wharf, Birkenhead, CH41 1LD, UK \\
$^{2}$Subaru Telescope, National Astronomical Observatory of Japan, 650
  North A'ohoku Place, Hilo, HI 96720, USA \\
$^{3}$Harvard College Observatory, 60 Garden Street, MS 42, Cambridge, 
  MA 01238, USA \\
$^{4}$Department of Physics and Astronomy, University of Leeds, LS2 9JT, UK \\
$^{5}$Cavendish Laboratory, J J Thompson Avenue, Cambridge, CB3 0HE UK }
\begin{document}

\date{Accepted 2007 ???. Received 2007 ??? ; in original form 2007 ??? }

\pagerange{\pageref{firstpage}--\pageref{lastpage}} \pubyear{0000}

\maketitle

\label{firstpage}

\begin{abstract}
We have surveyed a $\sim$0.9-square-degree area of the W\,3 giant molecular
cloud and star-forming region in the 850-$\umu$m continuum, using the SCUBA
bolometer array on the James Clerk Maxwell Telescope.  A complete sample
of 316 dense clumps was detected with a mass range from around
13 to 2500\,\msun.  Part of the W\,3 GMC is subject
to an interaction with the H{\sc ii} region and fast stellar winds generated 
by the nearby W\,4 OB association.  We find that the fraction of total gas 
mass in dense, 850-$\umu$m traced structures
is significantly altered by this interaction, being
around 5\% to 13\% in the undisturbed cloud but $\sim$25 -- 37\% in the 
feedback-affected region.  
The mass distribution in the detected clump sample depends somewhat 
on assumptions of dust temperature and
is not a simple, single power law but contains significant structure
at intermediate masses.  
This structure is likely to be due to crowding of sources near or below the
spatial resolution of the observations.
There is little evidence of any difference between 
the index of the high-mass end of the clump mass function in the 
compressed region and in the unaffected cloud. 
The consequences of these results are discussed in
terms of current models of triggered star formation.
\end{abstract}


\section{Introduction}

The most general observable quantities in star-forming regions that
can be related to predictive models are the star-formation efficiency
(SFE) and the initial mass function (IMF).  The average SFE is generally low
($\le 1$\%, Duerr, Imhoff \& Lada 1982) in molecular clouds in the Galaxy and
in normal external galaxies but can increase dramatically (by up to
$\sim 50$ times) in starburst galaxies (Sanders et al.\ 1991) and
galaxy mergers, an effect which has been linked to strong feedback
and enhancements in average gas density (e.g.\ Rownd \& Young 1999).

The origin of the stellar IMF is not yet clear, but one possibility
is that it is directly physically related to the rather similar mass function
of the dense clumps that are formed in the turbulent environment of star forming
regions (e.g.\ Clarke 1998 and references therein; Nutter \& Ward-Thompson
2007).  If so, then the stellar
IMF is determined by the physics of star formation and the basic nature of 
molecular clouds.  While the observed mass function in dense clumps is somewhat
variable from region to region (e.g. Johnstone et al.\ 2000; 2001),
there is little strong evidence of significant
variations in the stellar IMF (Massey 2003).
%(but see Weidner C \& Kroupa P 2005 ApJ 625 754 for extragalactic case)

\begin{figure*}
\includegraphics{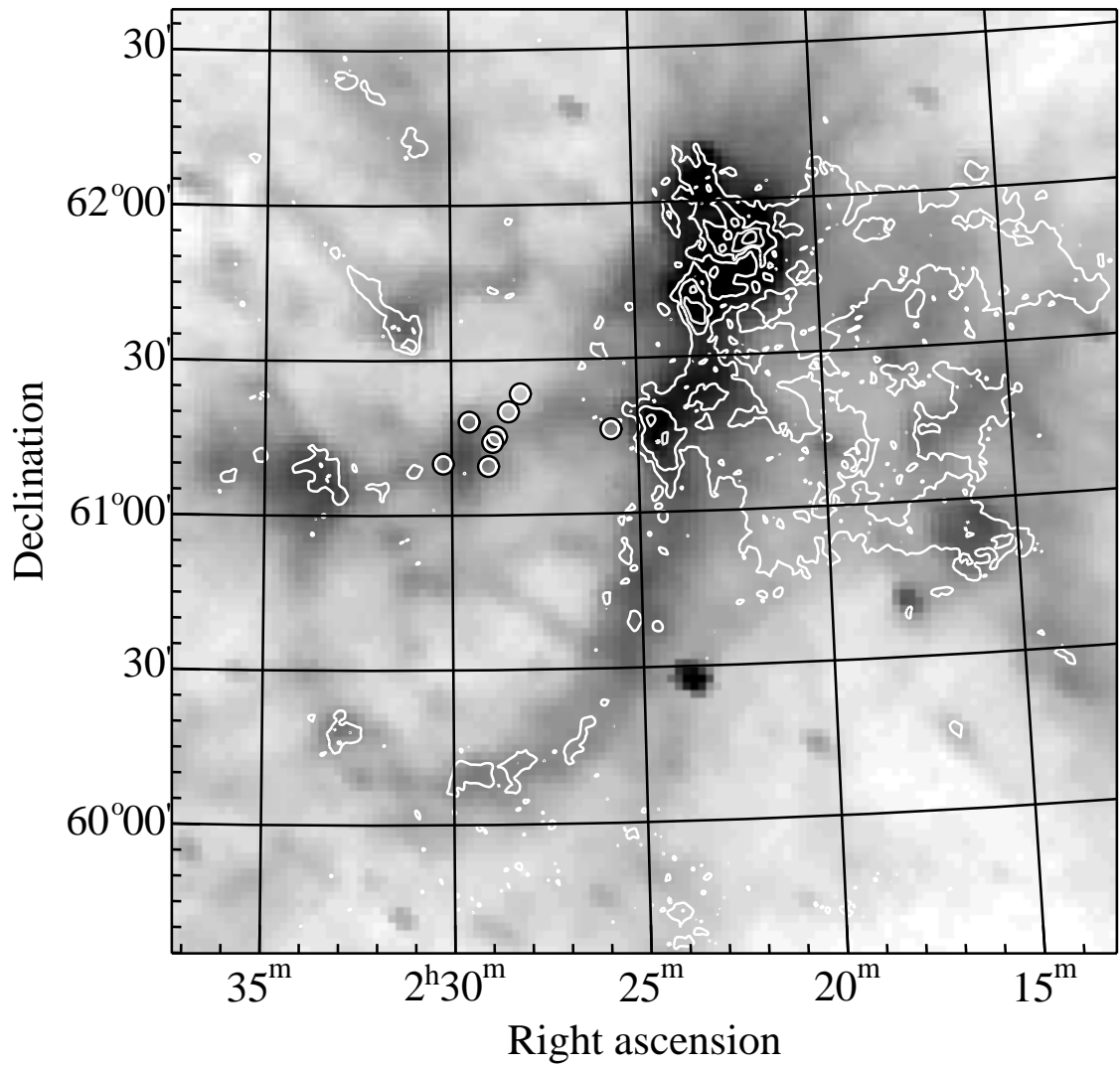}
  \vspace*{15cm}
  \caption{
Overview of the W\,3 GMC and the area immediately to the east.  The grey scale
shows MSX 8-$\umu$m emission delineating the edge of the W\,4 H{\sc ii} region
and the luminous star-formation regions in the eastern layer of W\,3.  The
white contours are integrated $^{12}$CO J=1--0 emission from the FCRAO Outer
Galaxy Survey (Heyer et al.\ 1998).  The positions of the O stars of the 
IC\,1805 cluster (Massey et al.\ 1995) are shown as circles.
}
\end{figure*}

Turbulent fragmentation models of star formation (e.g.\ Padoan \& Nordlund
2002) predict that complete Salpeter-like mass functions of
gravitationally bound dense clumps will form spontanously in molecular
clouds with driven turbulence.  Such models also suggest that the SFE is
determined by a combination of the scale on which the turbulence is driven
and the Mach number of the driven turbulence (e.g. V\'azquez-Semadeni et
al.\ 2003).  This is the paradigm of {\em spontaneous} star formation.

Investigating the physical basis for the idea of {\em triggered} star formation,
Whitworth et al.\ (1994) modelled the shocked, compressed cloud layers formed
by interactions.  They concluded that the dynamical
instabilities in the shocked gas generate new density structure, some of
which subsequently collapses, and predicted that higher-mass stars should be
preferentially formed under these conditions.
Lim, Falle \& Hartquist (2005) modelled the evolution of a cloud containing
a significant magnetic field subject to a sudden increase in external
pressure.  Their simulation predicts the formation of new dense structures, 
in which the thermal and magnetic pressures are comparable, which are 
potential sites of high-mass star formation.  

\begin{figure*}
%\special{psfile=fig1v1bw.ps
%  hoffset=-10 voffset=-590 hscale=90 vscale=90}
\includegraphics{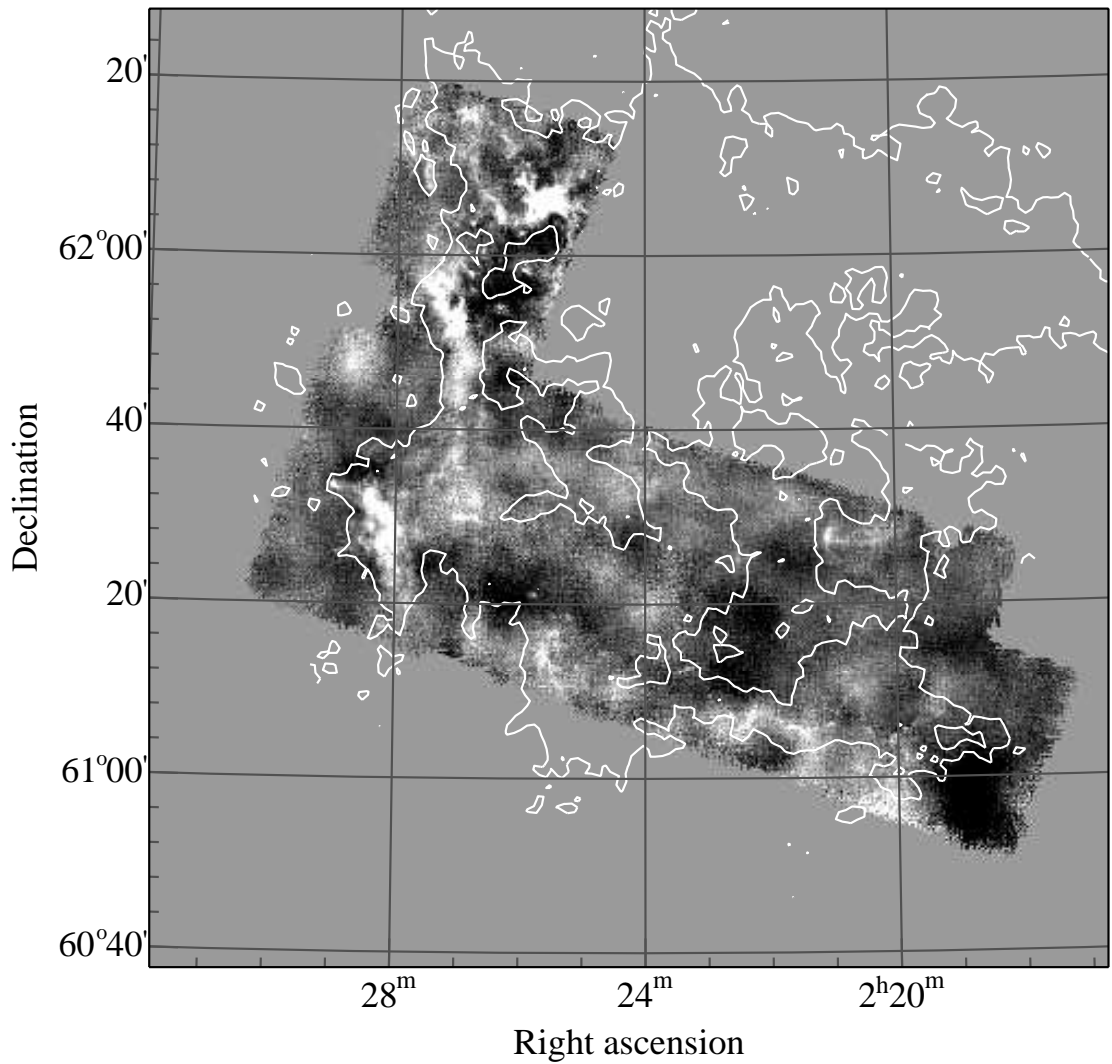}
  \vspace*{15cm}
  \caption{
The reduced but unflattened SCUBA 850-$\umu$m continuum emission map 
in grey scale, showing the surveyed portion of the W\,3 GMC and the 
large-scale signal artifacts arising from the observing technique.
The white contour follows the low-level integrated $^{12}$CO J=1--0 emission 
at 50$''$ resolution, from Bretherton (2003),
tracing the location and extent of the molecular gas in the W\,3 GMC, 
}
\end{figure*}

Surveys of Giant Molecular Clouds (GMCs), the sites of cluster formation, 
provide the observational constraints within which theoretical models must
operate. Thermal emission from the cold dust present in molecular
clouds peaks in the submillimetre, and is a reliably optically
thin tracer of column density.  Hence observations in the sub-millimetre
continuum are ideally suited to locating and quantifying the dense structure
which contains the current and incipient star formation (e.g.\ 
Johnstone et al.\ 2000).

The W\,3 GMC is a high-mass star-forming region located in the outer
Galaxy, in the Perseus spiral arm, at $l \simeq 134^{\rm o}$ and a distance 
of $\sim$2.0 kpc (e.g.\ Hachisuka et al.\ 2006) from the Sun.  The cloud 
occupies a well defined $1.5 \times 1.5$ degree area
and is one of the most massive molecular
clouds in the outer Galaxy (Heyer \& Terebey 1998).

Figure 1 shows the location of the W\,3 cloud, its 
proximity to the IC\,1805 OB association, the boundary of the
W\,4 H{\sc ii} region and the location of the high-mass star formation 
within the cloud itself as traced at 8$\umu$m by MSX.

Approximately 40\% of the cloud's total mass (Allsopp et al., 2007) is 
located in a layer of strong CO emission referred to as the high-density 
layer (HDL) by Lada et al.\ (1978).  The HDL occupies the eastern edge of
the GMC and runs parallel to the edge of the W\,4 \hii\ region. It is likely 
to have formed from compression of the cloud gas resulting from the
expansion of the \hii\ region and/or the ram pressure from the fast stellar
winds from the W\,4 OB association.  The luminous, massive star-forming 
regions within the HDL, (W\,3\,Main, W\,3\,(OH) and AFGL\,333), 
are likely to be examples of triggered star formation. 
The rest of the cloud is apparently unaffected by this or any other
major interaction with the surrounding medium.  The W\,3 GMC thus provides
a useful specimen for studying the differences between
induced and spontaneous star formation independent of initial conditions.

This paper presents the results of a census of star-formation activity
and dense structure in the W\,3 GMC, made to test the predictions
of models such as those mentioned above and to look for differences in the
mass distribution and fractional mass in dense structures that can
be related to the spatial variation in local conditions.

\section[]{Observations and data reduction}

The observations were made during 2001 July 17--24 and 26,
using the Submillimetre Common-User Bolometer Array (SCUBA) receiver
(Holland et al.\ 1999) on the 15-m James Clerk Maxwell
Telescope (JCMT). SCUBA is a dual-camera system, with 
91 bolometers optimized for performance at 450\,$\umu$m and 
37 bolometers for 850\,$\umu$m.  The spatial resolution at 450\,$\umu$m 
is 8 arcsec at FWHM and at 850\,$\umu$m it is 14 arcsec.
The two arrays observe simultaneously but the atmospheric opacity
at the time of the observations was too high for photometric 450-$\umu$m
data to be obtained.  Consequently, only the 850-$\umu$m data are
considered in what follows.
The SCUBA field of view is $\sim$2.3 arcmin
across. An area of 3150 arcmin$^2$, encompassing the HDL and the
southern portion of the W\,3 GMC, was surveyed (Figure 2).

Since the W\,3 GMC is considerably larger than the array field of view,
the data were obtained in scan-mapping mode using the ``Emerson II''
observing technique. A differential map of the source was generated by
% reconstruction Emerson2 technique \citep{1995mfsr.conf.....E}
scanning the array across individual 10-arcmin square fields
whilst the secondary mirror chopped in right ascension or declination
by one of three small chop throws -- 30 arcsec, 44 arcsec or 68 arcsec. 
%Chopping
%removes the DC offset due to sky emission and diminishes the effect of
%sky variability on the signal \citep{2000adass...9..559J}.  The
%resultant image consists of the sky distribution convolved with the
%primary beam and a dual-beam function of positive and negative delta
%functions separated by the chop throw
%\citep{2000ApJ...545L.121P,1979A&A....76...92E}.
Thus each 10-arcmin square submap comprises six component maps
of three chop throws in two directions. % In total, one hundred and
%ninety-two 10\am\ $\times$ 10\am\ individual chop-maps were acquired.

Pointing observations were made toward W\,3\,(OH) every
$\sim$90 minutes, and found to vary by less than 6 arcsec in Right
Ascension and Declination. The zenith atmospheric opacity was estimated by
performing skydips approximately every hour.  The average atmospheric
opacity at 850\,$\umu$m was 0.356, whilst the maximum and minimum values were
determined to be 0.702 and 0.185 respectively.

Data reduction was done using the software package {\sc surf}
(Jenness \& Lightfoot 1998). 
The data were flat-fielded, extinction corrected and despiked. 
Noisy bolometers were identified and blanked. Due to the difficulty
inherent in identifying emission-free regions for the purposes of
signal baseline removal, the median level was removed from each scan
(Johnstone et al.\ 2000). 
A model of the sky atmospheric emission
was calculated and removed from the scan-map data. 

The data were flux calibrated using observations of Uranus and the
planetary nebula CRL\,618. The Starlink program {\sc fluxes}
%(Privett et al.\ 1998?) 
was used to calculate flux densities for Uranus,
whilst the flux density of CRL\,618 was assumed
to be $4.56 \pm 0.17$\,Jy at 850\,$\umu$m. The derived flux conversion
factors were applied to the individual chop-maps prior to rebinning.
Calibration is estimated to be accurate to 6\%, based on the root mean
square deviation of the flux conversions factors.
The rebinned chop-maps were then mosaiced together with other maps
taken with the same chop configuration. The resulting six large maps
were Fourier-transformed and combined in Fourier space; a filter was
applied to remove data at spatial frequencies above SCUBA's sensitivity
threshold (ie structure smaller than the beam).
An inverse FT then generated the reconstructed map (Figure 2).

\subsection{Suppression of extended structure}

\begin{figure*}
\includegraphics{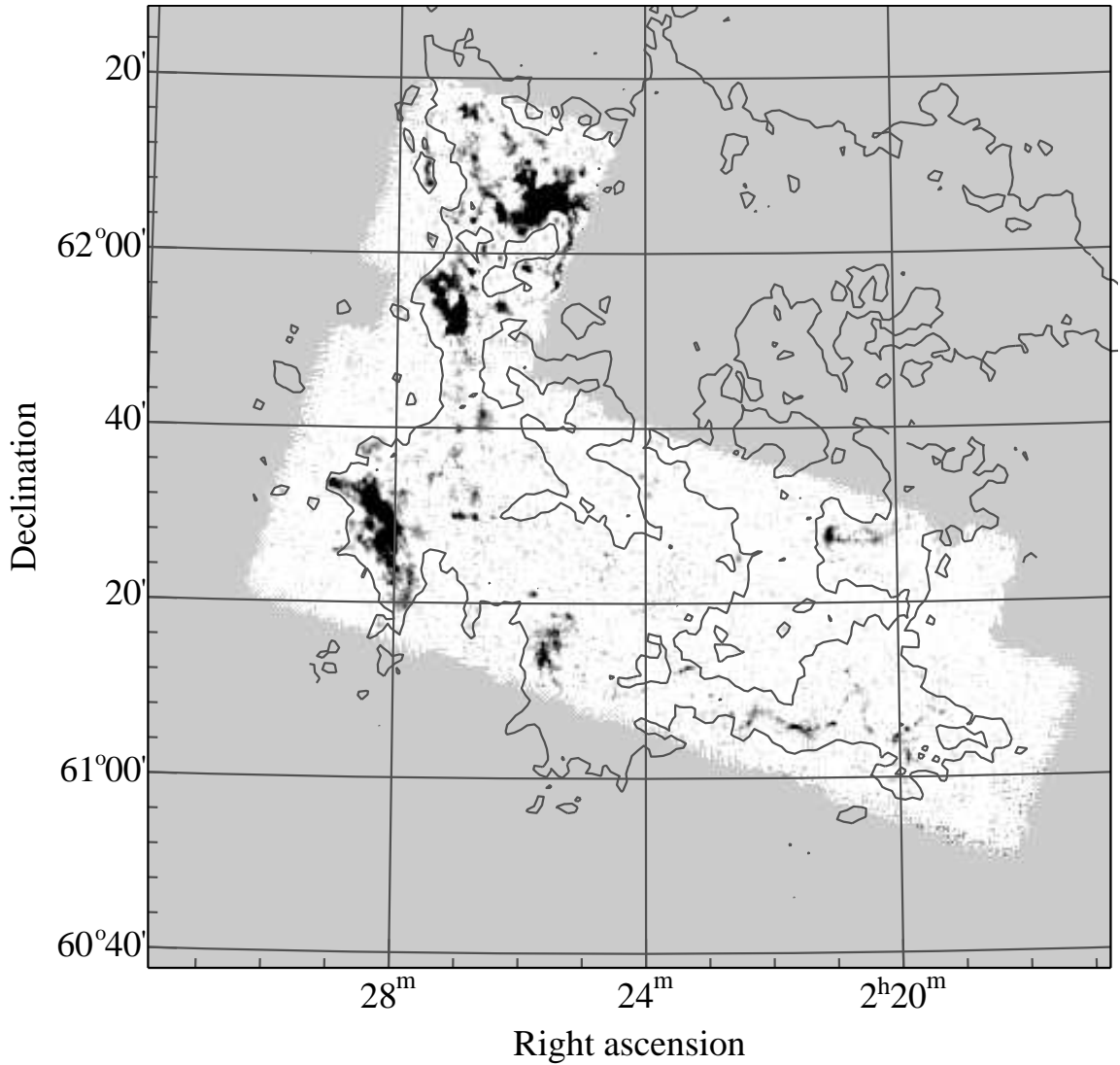}
  \vspace*{15cm}
  \caption{The SCUBA 850-$\umu$m continuum map after flattening (see text).
The continuum emission is shown in negative logarithmic grey scale and the cloud
boundary is marked in grey contours, as in Figure 2.
}
\end{figure*}

In SCUBA's scan-mapping mode, spatial scales more extended than a few
times the maximum chop throw are measured with significantly 
reduced sensitivity.
However, the observing and map reconstruction methods described above 
produced 
%two types of apparent extended structure in the reduced maps.  The first was 
semi-periodic apparent structure on scales similar to the size of 
the individual scan maps (10 arcmins) with amplitude $\sim 0.15 - 0.2$\,Jy
per pixel (4 to 5 times the unsmoothed pixel-pixel rms noise; see below).

In order to suppress this spurious extended structure, a template was
constructed by subtracting and interpolating over all strong steep-gradient 
sources from the reduced map and smoothing the result with a Gaussian 
function of FWHM 80\,arcsec (a little larger than the maximum chop throw).  
This template was subtracted from the original reduced map and the
resulting image was smoothed to the resolution of the telescope (14 arcsec)
to yield the final processed map (Figure 3). 
Removal of the extended structure from the map is largely cosmetic and 
introduces additional flux
measurement uncertainties and necessarily deletes a certain amount
of real low-frequency structure.  It also tends to produce slight 
negative `dishing' around the brightest sources.  The latter is difficult
to avoid in the more crowded regions where merged sources and the artifacts
become difficult to distinguish, and sources are difficult to remove
accurately in the creation of the background template.

The mean pixel-to-pixel rms noise in the processed, smoothed map 
was found to be $\sim$13\,mJy, which is equivalent to 56\,mJy per beam.

%(35\,mJy/beam in unsmoothed -- 
%x 4.2 = 150\,mJy per beam if 17.7 pixels per beam. This gives 3-sigma
%detection limit of 440\,mJy [or 160\,mJy if can use the smoothed rms value]).

\section{Results and analysis}
\subsection{General features of the processed map}

Figure 3 shows that the strongest
850-$\umu$m emission is located in the HDL region along the eastern
edge of the cloud. The conspicuously bright sources in the northern part
of this strip correspond to the known star-formation regions W\,3\,Main 
%(at $l = 133.71^{\circ}$, $b = 1.21^{\circ}$) and W\,3\,(OH) 
%($l=133.95^{\circ}$, $b=1.06^{\circ}$).  
(at {\sc ra} $\simeq 2^{\rm h}25^{\rm m}38^{\rm s}$, {\sc dec} $\simeq
+62^{\circ}05^{'}58^{''}$) and W\,3\,(OH) 
%($l=133.95^{\circ}$, $b=1.06^{\circ}$).  
($2^{\rm h}27^{\rm m}05^{\rm s}$, $+61^{\circ}52^{'}05^{''}$).
There is a considerable amount of strong emission in 
the environs of these two sources.
The W\,3\,North star-forming region is also distinguishable at 
%($l =133.78^{\circ}$, $b=1.42^{\circ}$). 
($2^{\rm h}26^{\rm m}54^{\rm s}$, $+62^{\circ}16^{'}06^{''}$).
Further to the south along the HDL, the AFGL\,333 region 
%($l=134.20^{\circ}$, $b=0.76^{\circ}$) 
($2^{\rm h}28^{\rm m}09^{\rm s}$, $+61^{\circ}30^{'}00^{''}$)
has a beaded, filamentary appearance. North east of AFGL\,333 there is 
a bright pointlike source corresponding to the position of the known 
outflow IC\,1805-W 
%($l=134.28^{\circ}$, $b=0.86^{\circ}$) 
($2^{\rm h}29^{\rm m}03^{\rm s}$, $+61^{\circ}33^{'}29^{''}$)
associated with IRAS point source 02252+6120.

In the south-western portion of the map the continuum emission is of
much lower average intensity. Many compact features are detected, however,
including a group of sources near 
%($l=134.02^{\circ}$, $b=0.40^{\circ}$) 
($2^{\rm h}25^{\rm m}39^{\rm s}$, $+61^{\circ}13^{'}34^{''}$)
and a sinuous filament which runs from 
%$l\sim133.7^{\circ}$ to $\sim133.5^{\circ}$. 
{\sc ra} $\sim 2^{\rm h}22^{\rm m}25^{\rm s}$ to 
$\sim 2^{\rm h}20^{\rm m}30^{\rm s}$, {\sc dec} $\sim +61^{\circ}06^{'}$.
Immediately west of this filament is a loop
of sources associated with, and possibly formed by the expansion of
the KR\,140 compact \hii\ region (Kerton et al.\ 2001).  North of KR\,140,
a prominent group of sources is located at 
%($l=133.43^{\circ}$, $b=0.43^{\circ}$).
($2^{\rm h}21^{\rm m}06^{\rm s}$, $+61^{\circ}27^{'}28^{''}$).

\subsection{Source detection}

Individual 850-$\umu$m sources were identified using
{\sc clfind2d}
%\footnote{Available
%  at {\it http://www.ifa.hawaii.edu/$\sim$jpw/research/ clfind/clfind.html}.},
 (version of 6/10/04) a two-dimensional adaptation of
the Williams, de Geus and Blitz (1994) clump-finding algorithm, {\sc clfind}.
This technique decomposes the data into a set of discrete clumps by
first (virtually) contouring the map at a series of levels set by
the user. The
clumps are then located by identifying emission peaks and tracing closed
contours down to lower intensities. Williams et al.\ (1994)
provide a detailed description of the algorithm's methodology and
performance testing with simulated data. The advantages of {\sc
  clfind} are that it does not assume any {\it a priori}
source profile and that it is an objective technique. Its weaknesses are
mainly those associated with the separation of crowded objects and the
inclusion of spurious sources at low levels, which are common to all
object-detection routines.

{\sc clfind} defines a clump boundary as 
the least significant contour surrounding the parent
emission peak. This boundary is signal-to-noise dependent and not
necessarily related to any physical outer radius.
This should introduce a bias against extended, low-surface-brightness
objects.  If the detection contour is set low so as to minimise this bias,
this will result in the overestimation of the fluxes of faint sources, 
relative to standard aperture photometry, since
source fluxes are integrated over all pixels within this boundary.

{\sc clfind2d} was run with input contour levels at 1$\sigma$ (13\,mJy), 
2$\sigma$
(27\,mJy) and 3$\sigma$ (40\,mJy) then at 3-$\sigma$ intervals up to the peak
signal of 14.2\,Jy per pixel.  The low sigma levels were included in order 
to ensure detection of faint but significant sources. With $\sim$18 pixels
inside the half-maximum radius of an unresolved source, a 5-sigma detection
may have a peak flux surface brightness that is barely above the 1-$\sigma$
level.  It also enables us to examine the spectrum of noise in the data and 
to measure the 
source detection and completeness limits, since these can not immediately
be inferred from the pixel-to-pixel noise.  In any case, adjacent pixels are
not independent because of the data reconstruction process. 
`Detections' less than 68 arcsec 
(the maximum chop throw) from the map edges were rejected since, in these
marginal regions, the residual noise is
greatest, coverage is incomplete and reconstructed fluxes are unreliable.

The distribution of the resulting sample, which is dominated by noise, 
is presented in Figure 4 as a 
histogram of equal-width flux bins.  Also shown in Figure 4 is the 
distribution of an equivalent sample obtained with the object 
detection routines in the Starlink {\sc gaia} package 
(which uses {\sc SExtractor}: Bertin \& Arnouts 1996). 
This extraction used elliptical isophotal fitting with the
same detection limit.  The two distributions are very 
similar except at very low flux levels well below the completeness limit.

\begin{figure}
\includegraphics{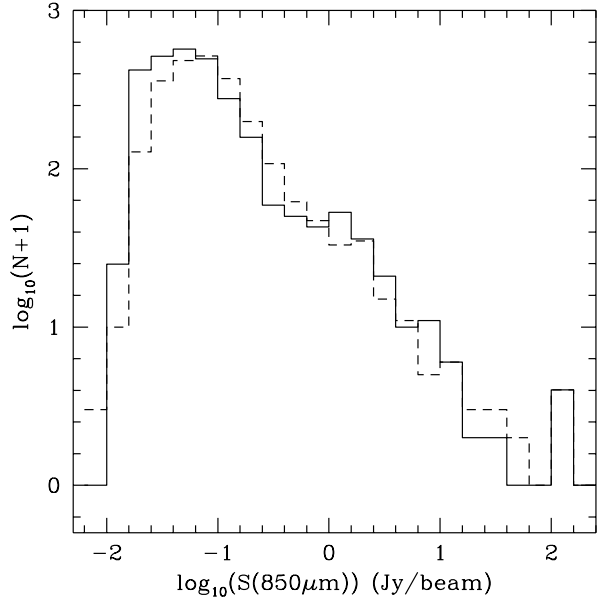}
  \vspace*{9cm}
  \caption{Distribution of the fluxes of all {\sc clfind2d} detections
above the 1-$\sigma$ per-pixel noise.  
The distribution is dominated by the noise spectrum at low flux levels.
The dashed histogram shows the results
obtained using {\sc gaia} object detection.
}
\end{figure}

\subsection{Completeness}

%The resulting list comprised 2741 apparent sources, the majority of which
%are noise. 

Figure 4 shows the combined noise spectrum and source flux distribution.  
It also shows that it is not
easy to distinguish the boundary between noise sources and real detections
in these data by simple inspection.

In order to define a completeness limit, we introduced 50 artificial
point sources, repeatedly and at random positions, into each of two 
otherwise blank regions of the reduced and flattened
image (Fig.\ 3) and used {\sc clfind2d} to recover them and measure fluxes.
By doing this, we find that the recovery rate is 100\% down to
flux densities of 113\,mJy per beam, dropping to 50\% at 28\,mJy per beam.  

The 1-sigma rms noise in the processed, smoothed map is 13.3\,mJy
per pixel or 56\,mJy per beam.
% should be right for a point source occupying 17.7 pixels, ie 13.333 x sqrt(17.7))}  
A completeness limit of 113\,mJy 
therefore corresponds to a 2-sigma detection and 
represents the total flux density of an unresolved source which
can be reliably recovered from the image using {\sc clfind2d}.  
Clearly, extended sources must have higher completeness limits.  

This is not
the end of the story, however, since we also need to know the flux level at
which all detected sources are real.  In other words, the limit of effectively
zero contamination by spurious noise sources.  By running {\sc clfind2d}
on the same two apparently empty regions of the image, we extracted a 
spectrum of detections resulting only from noise.  Scaling these
spectra by relative area to the total source flux distribution in 
Fig.\ 3, we find that contamination is significant ($>25$\% or around 1
detection per 10 square arcminutes) at flux densities
of 225\,mJy per beam and below, but effectively zero at 280\,mJy and above.  
The latter is 2.5 times the completeness limit obtained from the source 
recovery tests
and around five times the nominal 1-sigma noise (in mJy per beam). 
The latter flux density limit, which gives us a sample of 316 real sources, is
adopted in what follows.  The location of these sources within the 
survey area is plotted in Figure 5. 

%During this process we also found a tendency for the recovered fluxes of
%test objects to be spread somewhat
%by {\sc clfind2d}.  The observed tendency is for the fluxes of strong 
%sources to be underestimated and those of weak sources to be over-estimated.
%This creates a flattening of the flux distribution at low levels.  However,
%the effect is significant only at fluxes well below the completeness and
%contamination limits obtained above.

\begin{figure*}
\includegraphics{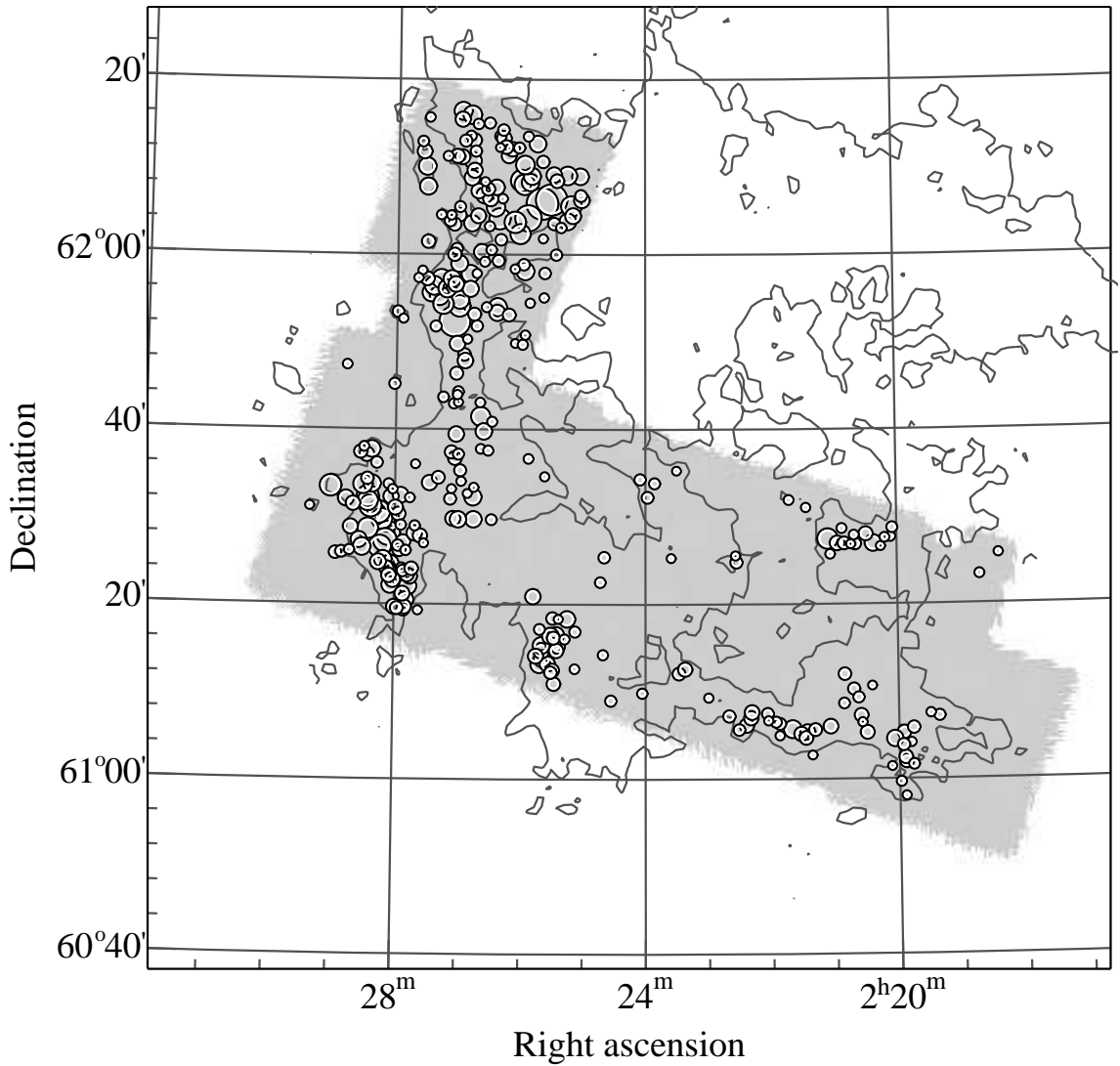}
  \vspace*{15cm}
  \caption{The SCUBA 850-$\umu$m source positions indicated by circles.  
The circle diameters are proportional to the log of the source flux density.
The continuum survey area is shaded in light grey and the cloud boundary, 
traced in $^{12}$CO J=1--0 as in Figure 2, is marked in dark grey contours.
}
\end{figure*}

\subsection{Dust temperatures and clump masses}

Under the assumption that the 850-$\umu$m emission is optically thin,
gas masses can be estimated from source flux densities $S_{\nu}$ using

\begin{equation}
M=\frac{S_{\rm \nu}D^{\rm 2}}{\kappa_{\rm \nu}B_{\rm \nu}(T_{\rm d})},
\end{equation}

\noindent
where $D$ the
assumed distance, {\mbox{$B_\mathrm{\nu}(T_\mathrm{d})$}} the Planck
function evaluated at dust temperature {\mbox{$T_\mathrm{d}$}} and
{\mbox{$\kappa_\mathrm{\nu}$}} is the mass absorption coefficient or
opacity.

In accordance with Mitchell et al.\ (2001), %\citet{2001ApJ...556..215M}, 
the value of the (gas plus dust) mass 
absorption coefficient at {\mbox{$\lambda=850\,\umu$m}} was taken to be
{\mbox{$\kappa_\mathrm{850}=0.01$\,cm$^\mathrm{2}$\,g$^\mathrm{-1}$}},
including an assumed gas-to-dust mass ratio of 100. Adopting a distance of
%2.5\,kpc to the W\,3 GMC (Digel et al.\ 1996), %\citep{1996ApJ...458..561D}, 
2.0\,kpc to the W\,3 GMC (Hachisuka et al.\ 2006), 
the above equation takes on the form

\begin{equation}
%M_{\rm clump}=46.26 S_{\rm 850} \left [ \exp \left ( \frac{17 K}{T_{\rm d}} \right ) -1 \right ] M_\odot
M_{\rm clump}=29.61\;S_{\rm 850} \left [ \exp \left ( \frac{17 K}{T_{\rm d}} \right ) -1 \right ] M_\odot
\end{equation}

\noindent
where $S_\mathrm{850}$, the total 850-$\umu$m flux density within the clump 
boundary, is measured in Janskys. 

Dust temperatures were estimated from gas kinetic temperatures obtained
from our own ammonia inversion-line measurements of a subsample of 44
clumps (Allsopp et al.\ 2007).  $T_{\rm d}$ values were assigned in two
ways, in order to investigate the effect on the measured clump mass function.
In the first, we assigned a single dust temperature to all sources, equal 
to the median NH$_3$ value in the measured subsample for the relevant cloud
region. These values were 18\,K for the HDL and 14\,K for the diffuse region.
Uncertainties in these temperature estimates can cause large errors in 
calculated masses. A 30\% uncertainty in the above values creates an error of 
around --30\%, +100\%, in calculated mass.
In the second method, measured NH$_3$ gas temperatures were
assigned to specific clumps, where available.  The rest of the sample was 
assigned temperatures
randomly from the set of NH$_3$ temperatures in the relevant section of the
cloud (Diffuse region or HDL).
Figure 6 shows the distribution of NH$_3$ temperatures used.

The properties of all 850-$\umu$m sources above the contamination limit 
are listed in Table 1. The source coordinates correspond to the peak 
flux positions and the total flux densities are without background
subtraction, since the latter is insignificant after the removal
of large scale structure in the map.

\begin{figure}
\includegraphics{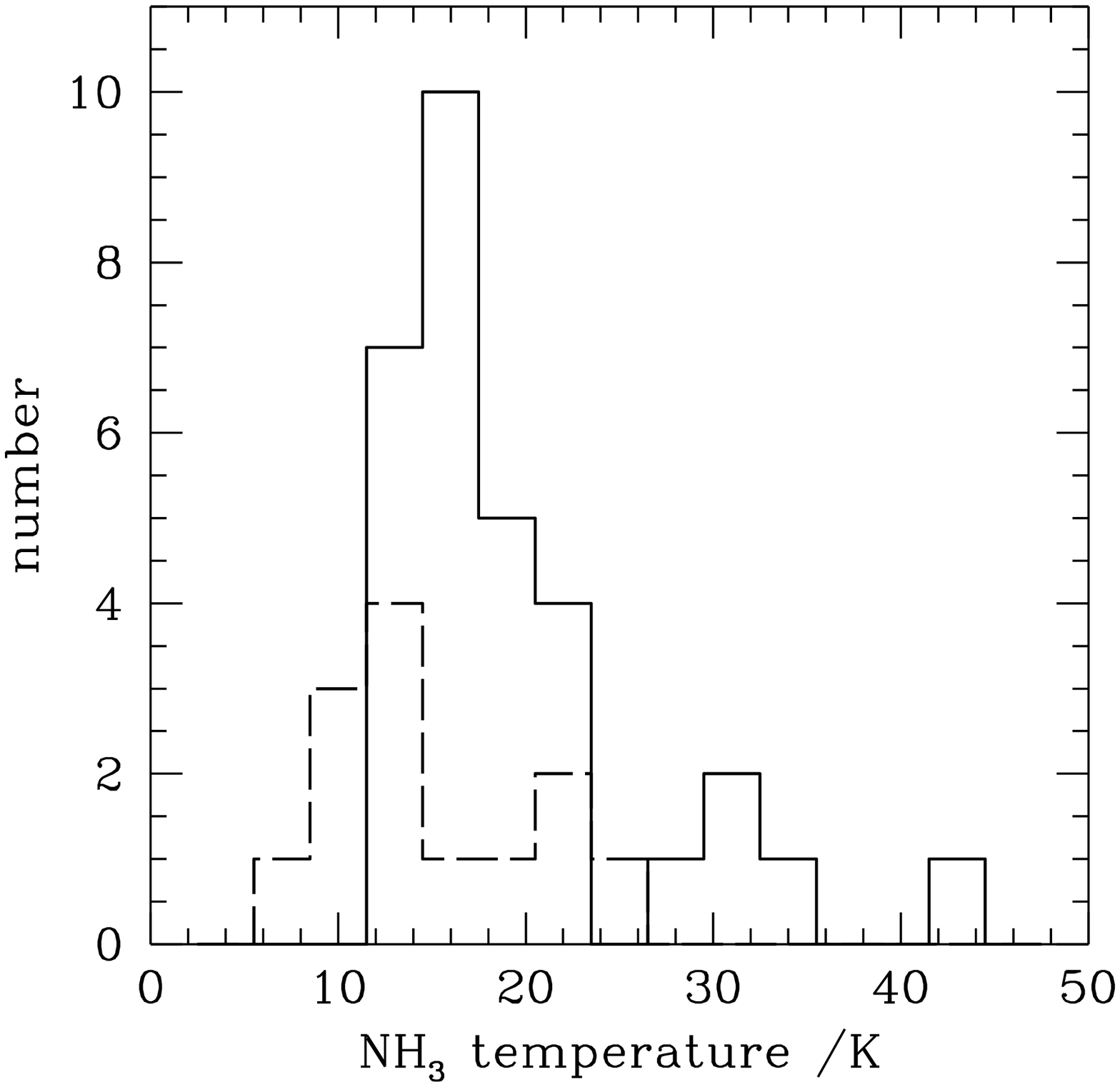}
  \vspace*{9cm}
  \caption{Distribution of NH$_3$ gas kinetic temperatures used in 
estimating clump dust temperatures.
The solid line denotes the HDL clump data, the dashed line is
the data for sources in the diffuse cloud.
}
\end{figure}

\subsection{The clump mass spectrum}

Figure 7 shows the distribution of clump masses, for sources with integrated
850-$\umu$m flux densities above the 280-mJy contamination limit, in both
the HDL and diffuse cloud region.  The masses entering these distributions
are calculated using both a fixed dust temperature and actual and 
randomly-assigned NH$_3$ temperatures (see above), averaged over 20 
repeats in the latter case.  These histograms have, as near as possible, 
equal population bins (and hence unequal bin widths) and are plotted as
the log of the bin population per unit log bin width.  The distributions in
the latter case turn over below $\log M \sim 1.1$ for the HDL clumps
and $\sim 1.3$ for the diffuse cloud data.  This is due to the 
combination of the flux completeness limit and the distribution of 
temperatures.

The single-temperature mass functions obviously follow the flux distribution
and are far from a single, simple power law in either of the two subsets.  
There is distinct structure around $\log M = 1.8$ in both subsamples.
This structure is still evident in the HDL data when 
actual and random temperatures from the NH$_3$ distribution 
are assigned to clumps (Figure 7b).  It does not appear in the diffuse cloud
sample in the latter model.

Above the completeness turnover masses, a single, linear fit to the averaged
HDL logarithmic mass function produced by the distributed-temperature model 
has a negative power-law index of $0.50 \pm 0.05$.  This fit includes the 
$\log M = 1.8$ structure 
and does not represent the data well.  The equivalent fit for the diffuse-cloud 
sample produces an index of $0.66 \pm 0.06$.  Above the apparent structure 
at $\log M = 1.8$, the HDL mass function is steeper (index $0.85\pm0.02$) 
and more consistent with a single power law.  The equivalent 
part of the diffuse-cloud data gives the same fitted index, within
the uncertainties, i.e.\ $0.80\pm0.06$.  Note that, in this form of the 
mass function, the canonical Salpeter stellar IMF has an index of 1.35.

In the single-temperature model (Figure 7a), a fit to all the data gives
results similar to those above.  Fitting only to data with 
$\log M > 1.8$ gives $0.92\pm0.06$ for the HDL sample index and $1.19\pm0.07$ 
for the diffuse cloud sample.  The former is consistent with the 
distributed-temperature model, but the latter is steeper, and rather 
closer to the Salpeter-like mass functions found in other studies.

If we consider only clumps with measured NH$_3$ gas temperatures, and use
these as dust-temperature estimates, the mass-function
fits have indices of $0.5\pm0.2$ for the HDL and $0.50\pm0.06$ for the 
diffuse cloud.  These are consistent with the fits to all data in both
temperature models, the larger error on the HDL result reflecting the 
persistent appearance of structure in the mass function in this subsample.

Many other determinations of clump mass functions use only those sources 
without evidence of star formation.  For consistency in the W\,3 
sample, all we
can do in this regard is remove the few clumps with IRAS and MSX 
Point Source catalogue detections.  There are 29 MSX point sources with
8-$\mu$m detections within 10$''$ of a SCUBA 850-$\umu$m source and a 
further 27 within 20$''$.  The former should be a reasonable association
criterion given the nominal pointing accuracy of MSX ($< 3''$; but see
Lumsden et al.\ 2002).  If the 10$''$ associations are removed from the
sample, using the single-temperature model and fitting to
$\log M > 1.8$, we get indices of $0.92 \pm 0.06$ and $1.5 \pm 0.1$ for 
the HDL and diffuse-cloud clumps, respectively.  The former is not
significantly different from the result using the whole sample, while
the latter is.  Removing the 20$''$ associations as well produces 
$0.94\pm0.07$ and $1.7\pm0.4$.

\begin{figure}
\includegraphics{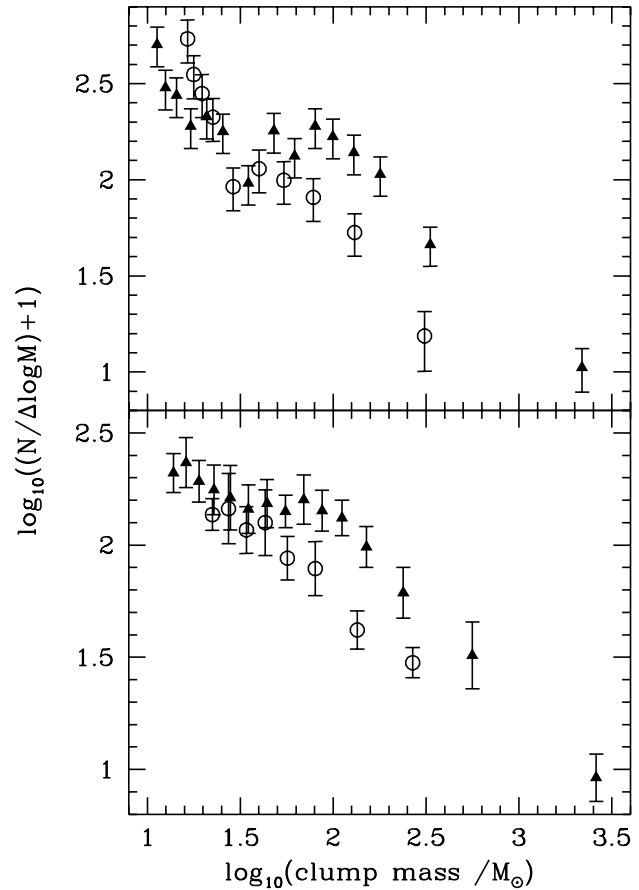}
  \vspace*{12.5cm}
  \caption{Distribution of masses of detected clumps above the noise 
contamination limit:  HDL (triangles) and Diffuse cloud (circles) 
samples.  Top: masses calculated using a single
dust temperature equal to the median of the gas temperatures in Figure 6
error bars are $\sqrt N$.
Bottom: using actual NH$_3$ gas temperatures, where available, 
otherwise a randomly assigned temperature from the distributions in
Fig 4 and is the average over 20 temperature assignments (errors are
standard deviations of the 20 results.
One has been added to mass bins to remove the possibility of taking the 
log of zero in an empty bin.
}
\end{figure}

%{\bf look at the CO data for outflows evidence - and this starless stuff 
%probably in a separate section to minimise confusion}

%A subset of the HDL sample from the Spitzer region alone gives a
%MF fit of $0.99 \pm 0.06$. The starless part of this subset gives $1.0 \pm 0.3$
%{\bf this is the only region in which it is meaningful to remove
%spitzer-selected star-forming cores}

\subsection{The fraction of gas mass in dense structures}

The total gas mass in the HDL and diffuse cloud regions was estimated from
maps of the whole W\,3 GMC in the $^{13}$CO 
%$^{13}$CO and C$^{18}$O 
J=1--0 rotational transition made at the FCRAO 14-m telescope (Allsopp et al.\
in preparation).  In order to calculate H$_2$ column densities, 
the LTE approximation was assumed with a single excitation temperature of 
30\,K.  The latter value is consistent with the colour temperature of the 
diffuse dust emission in IRAS extended emission maps (Bretherton 2003) and
adopting a higher temperature for the diffuse CO-traced gas than the dense
ammonia-traced clumps accounts for the likely greater penetration of the 
diffuse gas by radiation.  At temperatures above $\sim$10\,K, the column
density calculated from CO 1--0 is roughly proportional to the 
assumed exitation temperature. The probable error introduced by our 
assumption of 30\,K here is therefore no greater than other uncertainties
in calculating absolute column densities (see below).
%The rare isotopomer emission is used to estimate optical depths.  
The $^{13}$CO/H$_2$
%and $^{12}$CO/$^{13}$CO 
relative abundance was assumed to be $1.25 \times 10^{-6}$ and, for the 
purposes of this estimate only, the $^{13}$CO emission was assumed to be 
optically thin everywhere.
%and 80, respectively.

%(I get coadded 13CO in HDL to be 201668 Kkms/s and in diffuse cloud 323132.5 
%Kkms/s.  Values need correcting for pixels per beam (4?).  Areas were 11809 and
%33662 pixels, resp). 45" beam is 2.25e32 m2 at 2.5kpc.
%beam efficiency is about 50% and the spectral oversampling likely to 
%cancel this ~2?

Correcting for spatial oversampling and telescope beam efficiency, the 
total gas mass of the GMC was found to be $3.8 \times 10^5$\,\msun. 
%{\bf insert Jim's figures here}  
The mass in the HDL and the entire diffuse 
cloud region west of the HDL was $1.5 \times 10^5$ 
and $2.3 \times 10^5$\,\msun, respectively.  These estimates have a systematic
uncertainty which is a factor of order 2 -- 5 arising largely from the 
assumed relative abundances and CO excitation temperature.  
The corresponding CO-traced mass in the portion of the diffuse cloud region 
mapped with SCUBA is $1.15 \times 10^5$\,\msun.

The estimated total cloud mass gives an average gas density for the whole
GMC of only $4 \times 10^7$ m$^{-3}$, assuming the cloud is as deep along 
the line of sight as it is wide.  This is rather typical of GMC's and
means that the volume filling 
factor is $< 5$\%, if the CO-traced molecular gas density is above the
critical density for the 1--0 transition ($\sim 10^9$ m$^{-3}$ for $T \simeq 
30$\,K).

The total mass in dense clumps with 850-$\umu$m flux densities above the 
contamination limit of 280\,mJy per beam is $3.8 \times 10^4$ and $6.1 
\times 10^3$\,\msun\ in these two regions, respectively.  These figures 
assume the single-temperature model.  Using the temperature-distribution 
model the mass estimates are around 10\% higher.  The uncertainty in these
masses is dominated by errors in the assumed values of 
dust emissivity (see Henning et al.\ 1995) and 
dust temperature, and are a factor of order 2--3.  This uncertainty is,
again, largely systematic since we are dealing here with the sum over
the sample.  These mass values indicate that the detected fraction of gas
in the form of dense clumps is 0.26 in the HDL and 0.05
in the diffuse cloud.  These values are subject to the systematic
errors in the mass estimates, which combine into a factor of 4 or 5
but the difference between them is robust.

%In the Spitzer area of the HDL the fixed-T total core mass is 41675\,\msun.
%for the starless list, it's 6912 \msun.

\section{Discussion}

\subsection{General results}

The reduced 850-$\umu$m map of W\,3 (Figure 3) reveals that
the brightest sub-millimetre sources and a large fraction (86\% by mass;
69\% by number) of
the detected sources above the contamination limit of $\sim 13$\,\msun\ 
are located in the HDL.  This is not surprising, since the HDL contains the 
majority of the infrared sources associated with the cloud, several
well-known massive star-forming regions (W\,3\,IRS5, W\,3\,(OH), NGC\,333) each 
containing clusters
of (ultra) compact H {\sc ii} regions, and other phenomena associated with 
massive star formation, such as masers (e.g.\ OH and methanol: 
Etoka, Cohen \& Gray 2005) as well as energetic 
bipolar molecular outflows (e.g.\ Mitchell, Hasegawa \& Schella 1992).
This region of the cloud is apparently compressed by the expanding
H{\sc ii} region and/or the stellar winds from the W\,4 OB association 
and is likely to be the site of significant triggered star formation.  

Less predictably, we have found that there is a significant amount of dense
structure in regions of the W\,3 GMC which are far less affected by 
interactions.
14\% by mass and 31\% by number of the dense clumps were found in the portion
of the surveyed area away from the HDL (Figure 5).
The objects in this diffuse cloud region are all rather low mass, the 
brightest being
almost an order of magnitude fainter than the brightest source in the HDL
sample, but the lack of high-mass
clumps is consistent with statistics, as demonstrated by the similarity
in the high-mass slope of the mass function (see above).  The sources in 
the diffuse cloud are much less
densely packed than in the HDL.  It therefore appears 
that there is active, albeit
far less dramatic, star formation activity in the cloud away from the
regions where triggering by external interactions is dominant.  This star 
formation is likely to be associated with the natural turbulence in the cloud
as predicted by, e.g., Padoan \& Nordlund (2002), although
it is not possible to say that the diffuse cloud region is free of
external interactions.  The KR\,140 H{\sc ii} region at least
appears to have triggered the formation of a few small condensations by 
expanding into the southwest corner of the cloud (Kerton et al.\ 2001).  

Given that the observing technique is insensitive to emission that is 
extended on scales larger than the largest chop throw, 68 arcsec, 
we can only be concerned with the point-like sources in this study.
Despite this, there is some evidence in Figure 2 of a ridge of extended 
emission running through the middle of the HDL region, south from W\,3\,(OH).
This putative feature may have partly survived the method because it is 
rather narrow
in this region, but it has been further reduced by the large-scale background
removal that produces the final map (Fig.\ 3) and its significance is 
not considered further here.

The northern half of the western section of the cloud is not covered by
the present survey.  We can assume that it 
is similar in its physical state and star-formation content to the 
surveyed southern half but none of the conclusions we draw is dependent
on this assumption.  There is evidence of star formation activity in
this northern region (Bretherton 2003) and there have been suggestions that
this part of the cloud interacts with the HB3 (G\,132.6+1.5) supernova remnant 
(Routledge et al.\ 1991).  
%On the other hand, there is no evidence in this
%region of the increased velocities in the CO-traced gas, which we might
%expect to signify interaction with a fast shock (Allsopp et al.\ 2007).  

\subsection{The clump sample}

These observations give us a complete census of the dense, potentially
star-forming structures in the W\,3 GMC.  We have used the resulting sample
to construct the two basic quantities which define the star-forming
content: the distribution of masses (mass function) and the fraction
of the total cloud mass in dense structures that
may produce stars.  The fact that this particular GMC is experiencing
a major feedback interaction which affects only a well-defined section 
of the cloud
has allowed us to look for quantitative differences in these two parameters
that we might relate to the effects of the interaction.

The conservative limit to the sample reliability is the noise-contamination
limit at 280\,mJy.  Given our dust temperature and other assumptions above,
this gives the sample a lower mass limit of around 13\,\msun.
% this is actually 11M for HDL and 16M for Dif, but 13 is representative.
The sample of Enoch et al.\ (2006) from Perseus overlaps this, 
extending from a completeness limit of 0.8\,\msun\ to around 30\,\msun.

Above this limit, we have detected 316 sources, including 15 out of 20 
of the relatively faint objects found in the southwest corner of the cloud 
by Kerton et al.\ (2001), who also used SCUBA scan-mapping at 850$\umu$m.  
However, the 850-$\umu$m fluxes we obtain using {\sc clfind2d} are
three times larger, on average, 
than the values Kerton et al.\ (2001) measured using photometry within 
polygonal apertures.  The 850-$\umu$m fluxes are correlated, but flux 
ratios vary between 1 and 5, with a standard deviation of 1.3. 
The systematic discrepancy is partly due to the different photometry method
(\S 3.2) which will have the greatest effect on the weakest sources.

%comments on clfind - esp clustered sources in crowded fields.

\subsection{The dense clump mass function}

\begin{figure}
\includegraphics{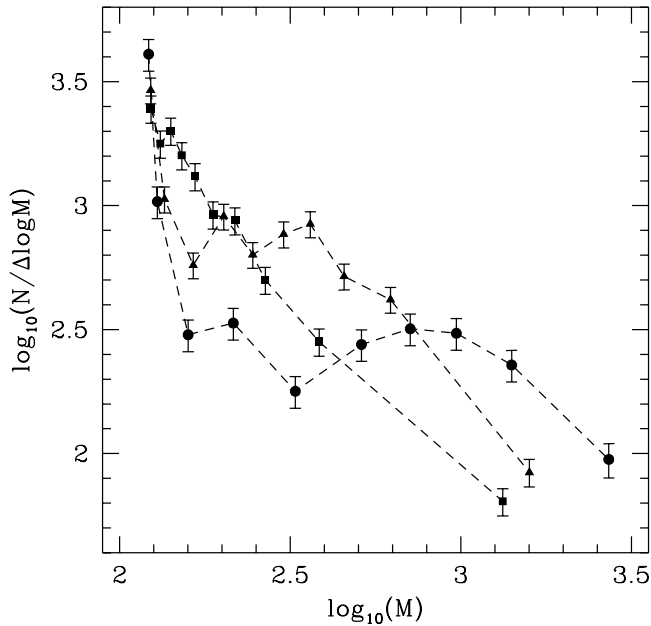}
  \vspace*{8.5cm}
  \caption{The results of using {\sc clfind2d} to recover a series of Monte 
Carlo 
simulations of sources with varying degrees of clustering.  Square symbols
represent a sequence with purely random positions. Triangles and
circles respectively represent sequences with Gaussian clustering probability
distributions having $\sigma \simeq 0.8$ and 0.5 times the spatial resolution. 
The dashed lines are added to clarify the separate sequences.
The mass scale is somewhat arbitrary, and results from the selected input
flux distribution (see text).
}
\end{figure}

The observed mass function in W\,3 is not a simple, single power law.  There
is evidence of a persistent feature, a slope reversal or peak, at masses of
around 60\,\msun\ (Figure 7).  The origin of this feature appears to be in
the combination of the scale on which sources are clustered (both 
physical clustering and superimpositions by chance along the line of sight) 
and the spatial resolution of the data.  
It should be noted that most, if not all, of the clumps detected in 
W\,3 are likely to form or be forming embedded clusters of stars.
All objects above the completeness limit are more massive than those in
Motte et al.\ (1998) that contain substructure.
The W\,3 clump mass function may, therefore, be more analagous to
a stellar cluster mass function.  

Where clustering causes small individual sources to merge so that they 
appear to be single, large objects, their mass will be counted in a higher-mass
bin.  In general, we might expect to find crowding depopulates the 
low- to intermediate-mass bins, since the most massive
sources will be rare enough that they are less likely to be merged together
or to have their fluxes significantly altered by the addition of smaller
objects.  

Figure 8 shows the result of a Monte-Carlo simulation of this
effect, generated as follows.  500 fluxes were extracted at random from 
a power-law distribution $N(F) \propto F^{-1.35}$ (matching the 
Salpeter IMF distribution) with limits 0.3 to 100\,Jy.  These
fluxes were assigned random positions within a source-free, 
$\sim 32 \times 10$ arcmins area of the 850-$\umu$m
map.  Where source clustering was required, ten seed fluxes were placed
in the field at random and the assigned positions of the rest were  
modified with a Gaussian probability distribution based on proximity to 
a previously placed object.  The resulting field, with its 500 artifical 
point sources, was smoothed to the resolution of the data (14.3 arcsec) and 
then the sources were recovered with {\sc clfind2d} exactly as for the
observed data.  This process was repeated 10 times for each simulation and
the resulting mass distributions (based on dust temperatures of 18\,K)
constructed.  Results of three different simulations are shown in Figure 8.  
These are for random positions (i.e.\ no clustering), for a Gaussian
clustering probability with standard deviation ($\sigma$)
equal to 3.9 pixels (12 arcsec or 0.8 times the spatial resolution), 
and for $\sigma = 2.2$ pixels (7 arcsec or about 0.5 times the spatial 
resolution).
% with exponent 1.20 (the generating mass function is actually the 
%Salpeter value of 1.35, but biases flatten the extracted MF).  
The results of this analysis are: 

\begin{itemize}
\item
The complex shape of the mass distribution measured in the HDL clumps
(Figure 7), including slope reversal and peak, is reproduced qualitatively when 
the clustering scale becomes comparable to the spatial resolution.  

\item
As the clustering becomes more severe, the position of the peak in
the mass distribution moves toward higher mass bins.

\item
The most likely origin of the observed reversal feature is in clustering
of sources at and below the spatial resolution, and not in any physical 
effect such as a second population of clumps with a lower-mass cutoff 
around 60 \msolar.  
\end{itemize}

A more detailed analysis of this effect may enable us to 
estimate the physical clustering at scales near and below the spatial 
resolution limit but this is beyond the scope of this paper.
In the mean time, it is interesting to point out that a just-resolved 
object with a mass equal to our sample limit (13\,\msun) 
would have a thermal ($T=20$\,K) Jeans length of $\sim 0.05$\,pc, close to the 
apparent clustering scale implied by the simulation.

As mentioned above, we might expect the highest mass clumps to be not 
significantly affected by crowding at the resolution scale.
%, since they
%are rare enough to be well separated and bright enough to remain dominant
%even when merged with many low-mass sources.  
We might thus expect to recover the original clump mass function index at the
high-mass end of the distribution.  Except where a single temperature is
applied, the fitted power laws to the upper end of the mass functions in
the two subsamples (\S 3.5) have exponents $\sim 0.84$, significantly 
flatter than that of a Salpeter IMF (1.35).  

Similarly flat mass distributions have been found in CO studies of cloud
structures (e.g.\ Williams et al.\ 2000), but those of the dense clumps 
traced by mm and submm wavelength data have generally yielded 
high-end power laws consistent with the Salpeter stellar IMF (e.g.\ Motte et al 
1998; Johnstone et al.\ 2000; Enoch et al.\ 2006) or steeper (e.g. 
Johnstone et al.\ 2001; Kirk et al.\ 2006).  The
reasons for the discrepancy between the molecular-line and (sub)mm continuum 
data are not clear.  There
has been much speculation in the literature of a direct connection
between these submillimetre clump mass functions and the stellar IMF via
turbulent fragmentation models of star formation.  The discrepant
results have not yet been fitted into such a picture but note that
determinations of {\em stellar cluster} mass functions yield power-law
exponents between 0.95 and 1.4 in our formulation, slightly flatter, on
average,
than Salpeter (Zhang \& Fall 1999, Lada \& Lada 2003, Hunter et al.\ 2003) 

It is worth noting that the observed mass function will tend towards the 
{\em cluster} mass function in the limit of strong clustering, and towards 
the clump mass function in the limit of zero clustering.  
Further, in the intermediate case, the existence of a cluster distribution 
similar to the clump 
mass function (ie with fewer high-mass clusters) will tend 
to steepen the observed clump spectrum (Weidner \& Kroupa 2005).  This occurs 
when the slope of the mass function of individual clumps is 
preserved within clusters.  Then many low-mass, and therefore truncated 
clump mass functions are superimposed 
on just a few high-mass clusters that extend over the whole mass range.

No clear difference has been found in the index at the high-mass end of the 
clump mass function (above 60\,\msun) between the HDL and the diffuse cloud
subsamples.  A difference does emerge when objects associated with MSX 8-$\mu$m
point sources are removed.  In this case the diffuse-cloud mass function
steepens into a power law consistent with Salpeter (\S 3.5).  This could 
mean that the fraction of clumps with embedded stars (and so
evolutionary status) is more mass dependent
in the diffuse cloud.  However, this result
must be treated with some caution as it may be the result of  
the MSX flux limit falling relatively high up in this lower-mass subsample.   
Any other differences between the two mass distributions
can be accounted for by the increased density of sources and the greater degree
of crowding in the HDL region.

The foregoing analysis, and the likelihood of unresolved clustering, 
shows that decoding the observed mass spectrum in W\,3 is a complex problem.
Until the spatial resolution available at these wavelengths is significantly 
improved, few strong constraints can be placed on the underlying distribution 
of clump masses, 
other than that they may be distributed as a power law with a negative exponent.

\subsection{The fraction of mass in dense clumps}

Using $^{12}$CO J=1--0 data, Lada et al.\ (1978) %\citet{1978ApJ...226L..39L} 
found the total mass of the W\,3 GMC to be
$\sim$7$\times$10$^{4}$\msolar.
They estimated masses of $\sim$4$\times$10$^{4}$\msolar\ and
$\sim$3$\times$10$^{4}$\msolar\ for the HDL and the diffuse cloud region
west of the HDL, respectively.  These are a factor of $\sim 5$ lower than
the estimates we use to calculate the mass fraction in dense clumps.  
This can probably be accounted for by optical depth effects,
choice of excitation temperature, undersampling in the older data, and
assumed abundance ratios.  Our figure of $\sim$$4\times10^{5}$\msun\ 
appears more consistent with the estimate of $\sim 10^6$\,\msun\ 
for the whole W\,3/4/5 GMC complex by Heyer \& Terebey (1998).  

The fraction of gas mass in dense, potentially star-forming structures detected
in these observations is around
26\% in the HDL region and only $\sim 5$\% in the diffuse cloud
area surveyed with SCUBA.  
These mass fractions are lower limits since there must be a component
of dense structures that is either extended and has been `resolved out' 
by the observing and reduction techniques and/or consists of compact 
sources below the detection limit.  The latter portion of this missing 
mass can be estimated by projecting the clump mass functions in Figure 7 
back to an assumed turnover 
mass of a little below 1 solar mass (e.g.\ Motte et al.\ 1998).  
The result of this depends on the exponent of the mass function at lower
masses.  Adopting a very flat power-law exponent of --0.5, consistent with 
the fit 
to the whole of the HDL sample, suggests that 8\% of the total mass in
dense clumps is undetected.  The equivalent missing fraction in the diffuse
cloud sample is 23\%.  This implies a corrected mass fraction in
dense clumps of 28\% in the HDL and 6.5\% in the diffuse cloud.  
If the exponent were as large as --1.5, 
%(consistent with the Salpeter stellar IMF and with the clump mass functions
%found by other authors at lower masses (e.g.\ Motte et al.\ 1998, +1)), 
these corrected mass fractions would rise to 37\% and 13\%, respectively.  

There is a large uncertainty (discussed above) in these absolute 
efficiency values, arising from adopted CO 
abundances and excitation, dust emissivity and temperature.  The two
mass fractions are, however, robust relative to each other.
We therefore conclude that there is a significant enhancement in the 
efficiency with which dense, potentially star-forming, structures are
formed from the cloud gas where that gas has been shocked by the
external interaction.  This enhancement is by a factor of at least 3
and possibly as high as 5.  

Since the HDL has apparently
been subject to a compressive interaction due to the expanding W\,4 \hii\
region, this result is consistent with either of two scenarios.  The first 
is that the effects of the interaction cause existing structures in the
cloud to accrete more material and grow more massive.  This may be due
to an increase in the signal speed in the compressed gas and, hence, in 
the accretion rate, or
an increase in the effective Jeans mass, both of which may be caused by
an increase in turbulent velocities.  The second possibility is that 
new dense structures are formed in the interaction, in the shocks between
turbulent flows or in local gravitational instabilities.
An increase in the fraction of total cloud mass contained in dense clumps 
is not consistent with a model in which feedback 
from previous generations of high-mass stars 
simply raises the ambient pressure and so increases the probability that 
existing structures collapse to form stars.  Feedback mechanisms must create
new dense structure from which stars can form or must force more of the 
cloud gas
into accretion flows onto existing bound objects.  This has a bearing on 
the question of how star formation efficiency is enhanced by feedback and 
is a clue to the origin of the 
large increase in star-formation efficiency observed in starburst galaxies,
for example.

This result is consistent with models of triggered star formation in which
entirely new structure forms as the result of an interaction (e.g.\ 
Whitworth et al., 1994, Lim et al., 2005).
It is also consistent with AMR simulations of the interaction of fast 
stellar winds
with turbulent clouds (Jones et al.\ in preparation).  These models
predict that density enhancements which form in the turbulent gas prior 
to the passage of the shock tend to continue to dominate in the post-shock 
gas.  The existing clumps are either stripped (if they are small) or 
accrete more material if they are massive and tightly bound.  This 
process might be expected to flatten the spontaneously 
formed dense clump mass function.  

\section{Conclusions}

We have surveyed two thirds of the area of the W\,3 Giant Molecular Cloud
in the 850-$\umu$m continuum at 14$''$ resolution, resulting in a complete
census of the star-formation activity in the surveyed region.  The 
observations produced a sample of 316 dense clumps above a flux limit
determined by contamination of spurious noise sources at 280\,mJy per beam.  
This limit is around five time the nominal 1-$\sigma$ noise level and
gives a lower mass limit of around 13\,\msun, depending on temperature
assumptions.  Analysis of the distribution of masses in the sample shows
that adopting a single temperature for all clumps produces a somewhat
different result from using a distribution of temperatures based on NH$_3$
gas temperatures.

The mass function is flatter than found in many other studies and is not
a simple, single power law but contains significant structure.  Simple
modelling indicates that this structure can be explained by crowding of
sources near or below the spatial resolution of the data.  Whether 
the implied characteristic scale ($\sim 0.1$\,pc) of this crowding 
is meaningful is not yet clear, but it is 
similar to the thermal Jeans length of just-resolved objects 
at the low-mass end of the sample.

The W\,3 GMC is subject to feedback from a previous generation of OB stars,
having been compressed on one side by the expansion of the H{\sc ii} region
generated by the W\,4 OB association, while the rest of the cloud is largely
unaffected.  The W\,3 cloud therefore provides a 
useful insight into the processes of triggered star formation and into the
differences between this and spontaneous star formation.

We have analysed the mass distribution and mass fraction in dense clumps
in the compressed region and the natural cloud.  There is little evidence
of any difference in the mass distribution, although more severe crowding 
in the compressed cloud layer may be having an effect.
The main difference comes in the fraction of the cloud
that has been converted to dense, potentially star-forming clumps.  This is
26 -- 37\% in the compressed region and only 5 -- 13\% in the 
diffuse cloud.  This difference suggests that the enhanced star-formation
efficiency associated with feedback and triggering is
not simply a process of increasing the probability that existing dense
clumps will collapse to form stars (e.g.\ by increasing the ambient pressure).
It is consistent with new structure being created
in the compressed shocked gas and also supports a model in which 
structures in the pre-shocked gas survive but accrete more efficiently 
in the post-shock environment.

\section*{Acknowledgments}

The James Clerk Maxwell Telescope is operated by The Joint Astronomy Centre 
on behalf of the Particle Physics and Astronomy Research Council of the 
United Kingdom, the Netherlands Organisation for Scientific Research, and 
the National Research Council of Canada.  DEB and JA acknowledge the support 
of PPARC studentships.

\begin{table*}
\begin{minipage}{130mm}
\begin{centering}
\caption{The 50 brightest sources from the W\,3 850-$\umu$m source sample.  The
full list of 316 objects is available in the on-line version}
\label{obstable}
\begin{tabular}{|l|l|l|l|l|l|}
ID$^a$ &  R.A. & Dec. & Peak S(850$\umu$m) & Integrated S(850$\mu$m) & IR associations \\ 
       &  (J2000)  & (J2000) & /Jy beam$^{-1}$ & /Jy & MSX$^{b,c}$ \& IRAS$^d$ \\
\hline
109 & 02$^{\rm h}\!$25$^{\!\rm m}\!$40$^{\rm s}\!\!$.2 & +62$^{\rm o}$05$'$49$''$ & $9.560\pm0.013$ & $198.9\pm0.7$ & G\,133.7150+1.2155$^b$ \\
    &              &           &      &       & IRAS\,02219+6152 \\
99 & 02 25 31.2 & +62 06 17 & 7.699 & $178.5\pm0.7$ & G\,133.6945+1.2166$^b$ \\
213 & 02 27 04.4 & +61 52 21 & 14.170 & $158.2\pm0.7$ & G\,133.9476+1.0648$^b$ \\
    &              &           &       &       & IRAS\,02232+6138 \\
119 & 02 25 53.8 & +62 04 10 & 2.774 & $41.6\pm0.7$ & \\
284 & 02 28 06.3 & +61 28 03 & 0.911 & $22.6\pm0.5$ & \\
288 & 02 28 09.1 & +61 27 14 & 0.743 & $19.38\pm0.4$ & \\
285 & 02 28 06.5 & +61 29 30 & 1.003 & $16.1\pm0.3$ & \\
292 & 02 28 15.4 & +61 30 29 & 0.616 & $14.3\pm0.4$ & IRAS\,02244+6117 \\
291 & 02 28 14.1 & +61 26 32 & 0.406 & $14.1\pm0.5$ & \\
297 & 02 28 23.4 & +61 31 10 & 0.545 & $13.3\pm0.4$ & \\
300 & 02 28 26.1 & +61 32 16 & 0.573 & $12.1\pm0.4$ & G\,134.2170+0.8135$^c$ \\
127 & 02 26 00.4 & +62 08 28 & 0.568 & $11.5\pm0.5$ & \\
315 & 02 29 02.0 & +61 33 26 & 0.927 & $11.0\pm0.5$ & G\,134.2792+0.8561$^b$ \\
    &            &           &       &       & IRAS\,02252+6120 \\
198 & 02 26 59.9 & +61 54 06 & 0.417 & $10.3\pm0.4$ & \\
132 & 02 26 06.5 & +62 03 42 & 0.319 & $10.3\pm0.5$ & \\
287 & 02 28 09.0 & +61 29 54 & 0.949 & $9.7\pm0.3$ & \\
142 & 02 26 21.5 & +62 04 06 & 0.393 & $9.6\pm0.4$ & IRAS\,02226+6150 \\
290 & 02 28 12.4 & +61 29 41 & 0.790 & $9.1\pm0.3$ & \\
124 & 02 25 57.4 & +62 08 01 & 0.320 & $8.5\pm0.5$ & \\
298 & 02 28 25.6 & +61 28 37 & 0.239 & $8.1\pm0.5$ & G\,134.2363+0.7539$^c$ \\
    &            &           &       &       & G\,134.2392+0.7511$^c$ \\
    &            &           &       &       & IRAS\,02245+6115 \\
78 & 02 25 12.0 & +62 05 32 & 0.201 & $7.9\pm0.4$ & \\
226 & 02 27 15.7 & +61 54 20 & 0.246 & $7.8\pm0.5$ & \\
38 & 02 21 06.0 & +61 27 28 & 0.302 & $7.7\pm0.5$ & IRAS\,02173+6113 \\
299 & 02 28 26.1 & +61 31 46 & 0.461 & $7.3\pm0.3$ & \\
128 & 02 26 01.4 & +62 02 21 & 0.490 & $7.0\pm0.5$ & \\
227 & 02 27 17.6 & +61 57 10 & 0.262 & $6.0\pm0.4$ & \\
230 & 02 27 19.6 & +61 55 16 & 0.423 & $5.8\pm0.3$ & G\,133.9598+1.1183$^b$ \\
304 & 02 28 30.9 & +61 33 39 & 0.174 & $5.8\pm0.4$ & G\,134.2176+0.8345$^c$ \\
163 & 02 26 38.5 & +61 41 28 & 0.164 & $5.5\pm0.5$ & \\
232 & 02 27 24.8 & +61 56 25 & 0.208 & $4.9\pm0.3$ & \\
123 & 02 25 57.1 & +62 10 16 & 0.217 & $4.6\pm0.4$ & \\
81 & 02 25 15.5 & +62 09 02 & 0.149 & $4.6\pm0.4$ & \\
181 & 02 26 49.0 & +62 15 58 & 0.275 & $4.5\pm0.4$ & G\,133.7836+1.4182$^c$ \\
    &            &           &       &      & IRAS\,02230+6202 \\
148 & 02 26 25.5 & +62 05 23 & 0.260 & $4.2\pm0.3$ & \\
234 & 02 27 27.2 & +61 55 37 & 0.293 & $4.1\pm0.3$ & IRAS\,02236+6142 \\
196 & 02 26 59.5 & +61 54 48 & 0.412 & $4.1\pm0.3$ & \\
145 & 02 26 23.0 & +61 53 59 & 0.281 & $4.1\pm0.3$ & G\,133.8572+1.0620$^c$ \\
223 & 02 27 12.0 & +61 56 05 & 0.171 & $4.0\pm0.3$ & \\
183 & 02 26 49.6 & +61 57 46 & 0.532 & $3.9\pm0.3$ & \\
192 & 02 26 58.1 & +62 16 24 & 0.320 & $3.9\pm0.3$ & \\
280 & 02 28 04.2 & +61 24 33 & 0.208 & $3.8\pm0.3$ & \\
45 & 02 21 40.8 & +61 05 42 & 0.345 & $3.8\pm0.3$ & \\
122 & 02 25 56.4 & +61 58 07 & 0.345 & $3.7\pm0.3$ & G\,133.7832+1.1065$^c$ \\
240 & 02 27 32.6 & +62 10 03 & 0.157 & $3.6\pm0.4$ & G\,133.8920+1.3601$^b$ \\
305 & 02 28 31.2 & +61 26 30 & 0.126 & $3.5\pm0.4$ & \\
289 & 02 28 10.9 & +61 25 06 & 0.173 & $3.4\pm0.3$ & \\
117 & 02 25 51.9 & +62 08 16 & 0.256 & $3.3\pm0.3$ & \\
220 & 02 27 07.4 & +61 57 11 & 0.163 & $3.2\pm0.3$ & \\
295 & 02 28 22.1 & +61 33 43 & 0.188 & $3.2\pm0.3$ & G\,134.2006+0.8304$^c$ \\
111 & 02 25 41.2 & +61 13 10 & 0.241 & $3.1\pm0.3$ & IRAS\,02219+6100\\
%103 & 02 25 37.9 & +61 13 49 & 0.240 & $3.1\pm0.3$ & G\,134.0223+0.3998$^c$ \\
%    &            &           &      &      & IRAS \\
%253 & 02 27 46.8 & +61 22 02 & 0.136 & $3.1\pm0.4$ & \\
%238 & 02 27 31.5 & +62 07 45 & 0.153 & $3.1\pm0.4$ & G\,133.9017+1.326$^c$ \\
%175 & 02 26 45.4 & +61 32 16 & 0.139 & $3.1\pm0.4$ & \\
%115 & 02 25 50.2 & +62 08 58 & 0.239 & $3.0\pm0.3$ & \\
%107 & 02 25 39.2 & +61 15 19 & 0.167 & $2.9\pm0.4$ & \\
%202 & 02 27 00.3 & +61 58 57 & 0.125 & $2.9\pm0.4$ & \\
%178 & 02 26 48.1 & +62 10 43 & 0.176 & $2.8\pm0.3$ & G\,133.808?+1.336$^b$ +IRAS \\
%80 & 02 25 15.3 & +62 03 47 & 0.161 & $2.7\pm0.3$ & \\
%296 & 02 28 22.6 & +61 37 40 & 0.092 & $2.7\pm0.4$ & \\
\hline
\end{tabular}
\end{centering}

$^a$Object number from full on-line source table\\
$^b$possible MSX association at separation $< 10''$\\
$^c$possible MSX association at separation $< 20''$\\
$^d$possible IRAS association at separation $< 60''$\\
\end{minipage}
\end{table*}

\bsp

\label{lastpage}

\end{document}